\documentstyle[prd,aps,twocolumn,epsf,floats,amsfonts,amssymb,amsmath]{revtex}

\newcommand{\be}{\begin{equation}}\newcommand{\ee}{\end{equation}}
\newcommand{\bea}{\begin{eqnarray}}\newcommand{\eea}{\end{eqnarray}}
\newcommand{\ba}{\begin{array}}\newcommand{\ea}{\end{array}}
\newcommand{\bit}{\begin{itemize}}\newcommand{\eit}{\end{itemize}}
\newcommand{\ben}{\begin{enumerate}}\newcommand{\een}{\end{enumerate}}

\newcommand{\lab}{\label}
\newcommand{\lf}{\left}
\newcommand{\noi}{\noindent}
\newcommand{\non}{\nonumber}
\newcommand{\ran}{\rangle}
\newcommand{\ri}{\right}

\newcommand{\al}{\alpha}\newcommand{\bt}{\beta}
\newcommand{\Ga}{\Gamma}
\newcommand{\De}{\Delta}
\newcommand{\te}{\theta}

\newcommand{\om}{\omega}

\newcommand{\ide}{1\hspace{-1mm}{\rm I}}
\begin{document}
\twocolumn[\hsize\textwidth\columnwidth\hsize\csname@twocolumnfalse\endcsname
%\preprint{Imperial/TP/97-98/79, hep-th/98}

\title{Quantization, group contraction and zero point energy}

\vspace{5mm}

\author {M. Blasone$^{1,4}$, E. Celeghini$^2$, P. Jizba$^3$  and
G. Vitiello$^4$\vspace{3mm}}

\address{$^1$ Blackett Laboratory, Imperial College, London SW7 1BZ, U. K. }
\address{$^2$ Dipartimento di Fisica, and Sezione INFN,
Universit\`a di Firenze, I-50125 Firenze, Italy}
\address{$^3$ Institute of Theoretical Physics, University of Tsukuba,
Ibaraki 305-8571, Japan}
\address{$^4$ Dipartimento di Fisica ``E.R. Caianiello'', INFN and INFM,
Universit\`a di Salerno, I-84100 Salerno, Italy }

\maketitle

\begin{abstract}
We study algebraic structures underlying 't Hooft's construction
relating classical systems with the quantum harmonic oscillator.
The role of group contraction is discussed. We propose the use of
$SU(1,1)$ for two reasons: because of the isomorphism between its
representation Hilbert space and that of the harmonic oscillator
and because zero point energy is implied by the representation
structure. Finally,  we also comment on the relation between
dissipation and quantization.
\end{abstract}

\vspace{8mm} ]

\section{Introduction}

Recently, the ``close relationship between quantum harmonic
oscillator (q.h.o.) and the classical particle moving along a
circle'' has been discussed \cite{thof1} in the frame of 't Hooft
conjecture \cite{thof3} according to which the dissipation of
information which would occur at a Planck scale in a regime of
completely deterministic dynamics would play a role in the quantum
mechanical nature of our world. In particular, 't Hooft has shown
that in a certain class of classical, deterministic systems, the
constraints imposed in order to provide a bounded from below
Hamiltonian, introduce information loss and lead to ``an apparent
quantization of the orbits which resemble the quantum structure
seen in the real world''.

Consistently with this scenario, it has been explicitly shown
\cite{BJV} that the dissipation term in the Hamiltonian for a
couple of classical damped-amplified oscillators
\cite{CRV,BGPV,BJ} is actually responsible for the zero point
energy in the quantum spectrum of the 1D linear harmonic
oscillator obtained after reduction. Such a dissipative term
manifests itself as a geometric phase and thus the appearance of
the zero point energy in the spectrum of q.h.o can be related with
non-trivial topological features of an underlying dissipative
dynamics.

The purpose of this paper is to further analyze the relationship
discussed in \cite{thof1} between the q.h.o. and the classical
particle system, with special reference to the algebraic aspects
of such a correspondence.

't Hooft's analysis, based on the $SU(2)$ structure, uses finite
dimensional Hilbert space techniques for the description of the
deterministic system under consideration. Then, in the continuum
limit, the Hilbert space becomes infinite dimensional, as it
should be to represent the q.h.o.. In our approach, we use the
$SU(1,1)$ structure where the Hilbert space is infinite
dimensional from the very beginning.

We show that the relation foreseen by 't Hooft between classical
and quantum systems, involves the group contraction \cite{inonu}
of both $SU(2)$ and $SU(1,1)$ to the common limit $h(1)$. The
group contraction completely clarifies the limit to the continuum
which, according to 't Hooft, leads to the quantum systems.

We then study the $D^+_k$ representation of $SU(1,1)$ and find
that it naturally provides the non-vanishing zero point energy
term. Due to the remarkable fact that  $h(1)$ and the $D^+_k$
representations share the same Hilbert space, we are able to find
a one-to-one mapping of the deterministic system represented by
the $D^+_{1/2}$ algebra and the q.h.o. algebra $h(1)$. Such a
mapping is realized without recourse to group contraction, instead
it is a non-linear realization similar to the Holstein-Primakoff
construction for $SU(2)$ \cite{holstein}.

Our treatment sheds some light on the relationship between the
dissipative character of the system Hamiltonian (formulated in the
two-mode $SU(1,1)$ representation) and the zero point energy of the
q.h.o., in accord with the conclusions presented in Ref.\cite{BJV}.

\section{'t Hooft's scenario}

As far as possible we will closely follow the presentation and the
notation of Ref. \cite{thof1}. We start by considering the
discrete translation group in time $T_1$. 't Hooft considers the
deterministic system consisting of a set of $N$ states, $\{(\nu)\}
\equiv \{(0), (1),...(N-1)\}$, on a circle, which may be
represented as vectors:
\bea\label{1}
(0)=\lf(\ba{c} 0 \\ 0\\ \vdots \\ 1\ea\ri) ;\, (1)=\lf(\ba{c} 1 \\ 0\\
\vdots \\ 0\ea\ri) ; \dots ; (N-1)=\lf(\ba{c} 0 \\ \vdots \\ 1\\ 0\ea\ri) ,
\eea
and $(0)\equiv (N)$. The time evolution takes place in discrete
time steps of equal size, $\De t=\tau$
\bea\label{1a}
t \to t + \tau \quad: \quad (\nu) \to (\nu + 1 ~~{\rm mod} ~N)
\eea
and thus is a finite dimensional representation $D_N(T_1)$ of the
above mentioned group. On the basis spanned by the states $(\nu)$,
the evolution operator is introduced as  \cite{thof1} (we use
$\hbar=1$):
\bea\label{2} U(\De t =\tau) \,=\, e^{-i H \tau} \, =\, e^{-i\frac{\pi}{N}}
\lf(\ba{ccccc} 0 &&&&1 \\ 1&0&&& \\ &1&0&&
\\ &&\ddots&\ddots& \\&&&1&0 \ea \ri)
\eea
This matrix satisfies the condition $U^N = \ide$ and it can be
diagonalized by a suitable transformation. The phase factor in
 Eq.(\ref{2}) is introduced by hand. It gives the $1/2$ term
contribution to the energy spectrum of the eigenstates of $H$
denoted by $|n\ran$, $n = 0,1,...N-1$:
\bea\label{3} &&
\frac{H}{\omega } ~~|n\ran ~~=~~ (n+\frac{1}{2})
~|n\ran ~, ~~~~ \omega \equiv \frac{2\pi}{N\tau} . \eea

The Hamiltonian $H$ in Eq.(\ref{3}) seems  to have the same
spectrum of the Hamiltonian of the harmonic oscillator. However it
is not so, since its eigenvalues have an upper bound implied by
the finite $N$ value (we have assumed a finite number of states).
Only in the continuum limit ($\tau \to 0$ and $l \to \infty$ with
$\omega$ fixed, see below) one will get a true correspondence with
the harmonic oscillator.

The system of Eq.(\ref{1}) is plotted in Fig.~1 for $N=7$. An
underlying continuous dynamics is introduced, where $x(t) = \cos
(\al t) \cos(\bt t)$ and $y(t) =  - \cos (\al t) \sin(\bt t)$. At
the times $t_j = j \pi/\al$, with $j$ integer, the trajectory
touches the external circle, i.e. $R^2(t_j)= x^2(t_j) +
y^2(t_j)=1$, and thus $\pi/\al$ is the frequency of the discrete
('t Hooft) system. At time $t_j$, the angle of $R(t_j)$ with the
positive x axis is given by: $ \te_j = j \pi - \bt t_j= j (1 -
\bt/\al)\pi$. When $\bt/\al$ is a rational number, of the form $q
= M/N$, the system returns to the origin  (modulo $2\pi$) after
$N$ steps. To ensure that the $N$ steps cover only one circle, we
have to impose $\al(t_j)= j \,2\pi/N$, which gives the condition
$M=N-2$. Thus, in order to reproduce 't Hooft's system for $N=7$,
as in Fig.~1, we choose $q=5/7$. For $N=8$, we have $q=3/4$ and so
on.
\begin{figure}[t]
%\vspace{-1cm}
\centerline{\hspace{0.6cm}\epsfysize=2.3truein \epsfbox{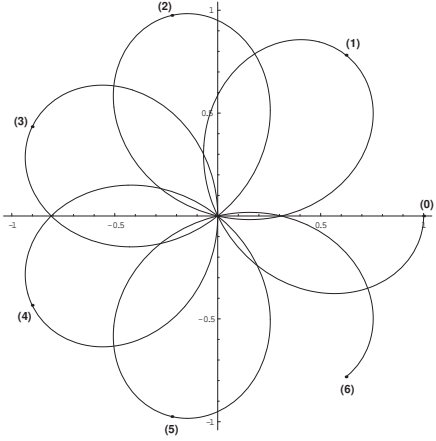}}
\vspace{0.5cm} \caption{'t Hooft's deterministic system for $N=7$.}
\vspace{0.3cm}
\hrule
\end{figure}

The system of Eq.(\ref{1}) can be described in terms of an
$SU(2)$ algebra if we set
\bea\label{4}
&&N \equiv 2l + 1 ~, ~~~n \equiv m + l ~, ~~~~m
\equiv -l,...,l ~, \eea
so that, by using the more familiar notation $|l,m\ran$ for the
states $|n\ran$ in Eq.(\ref{3}) and introducing the operators
$L_+$ and $L_-$ and $L_3$, we can write the set of equations
\bea \lab{5a}
\frac{H}{\omega} ~|l,m\ran &=& (n+\frac{1}{2}) ~|l,m\ran \,. \\
L_3 ~|l,m\ran &=& m ~|l,m\ran\,, \non \\
L_+ ~|l,m\ran &=& \sqrt{(2l-n)(n+1)} ~|l,m+1\ran\,, \non \\
L_- ~|l,m\ran &=& \sqrt{(2l-n+1) n}  ~|l, m-1\ran\,. \lab{5}
\eea
with the $su(2)$ algebra being satisfied $(L_{\pm} \equiv L_1 \pm
i L_2)$:
\bea \label{6}
&&[L_{i}, L_{j}\,] = i\epsilon_{ijk}L_k ~, ~~i,j,k
= 1,2,3 . \eea
't Hooft then introduces the analogues of position and momentum
operators:
\bea
{\hat x}\equiv \al L_x ,
\quad {\hat p}\equiv \bt
L_y ,  ~~ \al \equiv \sqrt{\frac{\tau}{\pi}} ,
~~ \bt \equiv \frac{-2}{2l +1} \sqrt{\frac{\pi}{\tau}}\, ,
\eea
satisfying the ``deformed'' commutation relations
\bea\lab{8}
[{\hat x},{\hat p} \,]\, =\, \al \bt i L_z \, =\, i\lf(1
- \frac{\tau}{\pi}H \ri)\,.
\eea

The Hamiltonian is then rewritten as
\bea\lab{9}
H \, =\,
\frac{1}{2} \om^2 {\hat x}^2 \, +\,\frac{1}{2} {\hat p}^2 \, +\,
\frac{\tau}{2 \pi}\lf(\frac{\om^2}{4} + H^2\ri)\,.
\eea
The continuum limit is obtained by letting $l \to \infty$ and
$\tau \to 0$ with $\omega$ fixed for those states for which the
energy stays limited. In such a limit the Hamitonian goes to the
one of the harmonic oscillator, the ${\hat x}$ and ${\hat p}$
commutator goes to the canonical one and the Weyl-Heisenberg
algebra $h(1)$ is obtained. In that limit the original state space
(finite $N$) changes becoming infinite dimensional. We remark that
for non-zero $\tau$ Eq.(\ref{8}) reminds the case of dissipative
systems where the commutation relations are time-dependent thus
making meaningless the canonical quantization procedure
\cite{CRV}.

We now show that the above limiting procedure
is nothing but a group contraction.
One may indeed define ~$a^\dagger \equiv
L_+/\sqrt{2 l}$,~ $a \equiv L_-/\sqrt{2 l}$~ and, for simplicity,
restore the $|n\ran$ notation ($n = m + l$) for the states:
\bea
\frac{H}{\omega} ~|n\ran &=& (n+\frac{1}{2}) ~|n\ran \lab{10a} \\
a^\dagger ~|n\ran &=& \sqrt{\frac{(2l-n)}{2 l}} \sqrt{n+1}
~|n+1\ran \,,  \non \\
a ~|n\ran &=& \sqrt{\frac{2l-n+1}{2 l}} \sqrt{n}  ~|n-1\ran\,. \lab{10}
\eea
The continuum limit is
then the contraction $l \to \infty$ (fixed $\omega$):
\bea
\frac{H}{\omega} ~|n\ran &=& (n+\frac{1}{2}) ~|n\ran \,.\lab{h1a} \\
a^\dagger ~|n\ran &=& \sqrt{n+1} ~|n+1\ran\,,  \non \\
a ~|n\ran &=& \sqrt{n}  ~|n-1\ran \,,\label{h1}
\eea
and, by inspection,
\bea
[a, a^\dagger ] ~|n\ran &=& ~|n\ran \lab{caa}\\
\{a^\dagger, a \} ~|n\ran &=& 2 (n+1/2) ~|n\ran \label{ca} .
\eea
We thus have $[a, a^\dagger ] = 1$ and $H/\omega = \frac{1}{2} \{
a^\dagger, a \}$ on the representation $ \{|n\ran\} $. With the
usual definition of $a$ and $a^\dag$, one obtains the canonical
commutation relations $[{\hat x}, {\hat p}]=i$ and the standard
Hamiltonian of the harmonic oscillator.

We note that the underlying Hilbert space, originally finite
dimensional, becomes infinite dimensional, under the contraction
limit. Then we are led to consider an alternative model where the
Hilbert space is not modified in the continuum limit.

\section{The $SU(1,1)$ systems}

The above model is not the only example one may find of a
deterministic system which reduces to the quantum harmonic
oscillator. For instance, we may consider deterministic systems
based on the non compact group $SU(1,1)$. An example is
the  system depicted in Fig.~2: It consists of
two subsystems, each of them made of a particle moving along a
circle in discrete equidistant jumps. Both particles and circle
radii might be different, the only common thing is that both
particles are synchronized in their jumps. We further assume that
for both particles the ratio (circumference)/(length of the
elementary jump) is an irrational number (generally different) so
that particles never come back into the original position after a
finite number of jumps. We shall label the corresponding states
(positions) as  $(n)_A$ and $(n)_B$  respectively. The plot in Fig.~2
is obtained by using the same continuous dynamics as for Fig.~1 with
$\bt/\al = 5/3 + \pi/40$.

The synchronized
time evolution is by discrete and identical time steps $\triangle
t = \tau$ as follows:
\begin{eqnarray*}
t \rightarrow t + \tau\,\,\,\, ; &&(1)_A\rightarrow (2)_A
\rightarrow (3)_A \rightarrow (4)_A \ldots\, , \\
&&(1)_B\rightarrow (2)_B \rightarrow (3)_B \rightarrow (4)_B \ldots\, .
\end{eqnarray*}

\begin{figure}[t]
%\vspace{-1cm}
\centerline{\hspace{0.6cm}\epsfysize=1.70truein
\epsfbox{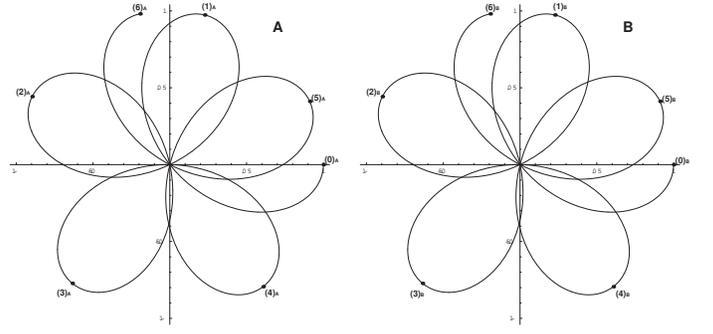}} \vspace{0.5cm} \caption{A deterministic
system based on $SU(1,1)$. } \vspace{0.3cm}
\hrule
\end{figure}

\noi This evolution is, of course, completely deterministic. A
practical realization of one of such particle subsystem is in fact
provided by a charged particle in the cylindrical magnetron, which
is a device with a radial, cylindrically symmetric electric field
that has in addition a perpendicular uniform magnetic field. Then
the particle trajectory is basically a cycloid  which is wrapped
around the center of the magnetron. The
actual parameters of the cycloid are specified by the Larmor
frequency $\omega_L = qB/2m$. To implement the discrete time
evolution we confine ourself only to an observation of the largest
radius positions of the particle. So we disregard any information
concerning the actual underlying trajectory. If the the Larmor
frequency and orbital frequency are incommensurable then the
particle proceeds via discrete time evolution with $\tau =
2\pi/\omega_L$ and returns into its initial position only after
infinitely many revolutions.

The actual states (positions) can be
represented by vectors similar in structure to the ones in
Eq.(\ref{1}) with the important difference that in the present case
the number $N$ of their components is infinite.
The
one--time--step evolution operator acts on $(n)_A\otimes (m)_B$
and in the representation space of the states it reads
\bea \non U(\tau ) \,&\equiv&\,
e^{-i H \tau} \, =\, e^{-i H_A \tau}\otimes e^{-iH_B
\tau}
\\
&=& \lf(\ba{cccc}  0&0&\ldots&1 \\ 1&0&\ldots&0
\\ 0&1&\ldots&0\\
&\ddots&\ddots  \ea \ri)_A \otimes \lf(\ba{cccc}  0&0&\ldots&1 \\
1&0&\ldots&0
\\ 0&1&\ldots&0\\
&\ddots&\ddots  \ea \ri)_B\, . \label{mat1}\eea
As customary one works with finite dimensional matrices and at the end of the
computations the infinite dimensional limit is considered.

It is worth to mentionthat our system consisting of two particles
``jumping" along two circles can, in fact, be realized with only a
single particle ``jumping" on a 2D torus. Assuming that
$\varphi_1$ and $\varphi_2$ are angular coordinates (longitude and
latitude) on the 2D torus we prescribe the one--time--step
evolution as
\bea \non t\rightarrow t + \tau\, ; \;\;\;\; &&\varphi_1 \rightarrow \varphi_1 +
\alpha_1 \tau\, ,\\&&\varphi_2 \rightarrow \varphi_2 + \alpha_2 \tau\, .
\eea
\noi Note that while the discrete time evolution on latitude
circle stays on the latitude circle, the discrete time evolution
along longitude does not preserve the longitude circle but deforms
it into ``winding line". This is not in contradiction with the
previous two--circle model.In reality, the (common) key point is
that after infinite (and only infinite) time the system returns
into the original position. In fact, in the torus system if
$\alpha_1/\alpha_2$ is irrational then the positions (states)
never return back into the original configuration at any finite
time but instead they fill up all the torus surface (they are
dense in the torus \cite{Arnold}).
 Inasmuch the states (positions) are dense along both ``circles"
separately and return into the initial position after infinitely
long time. We will not consider in this paper further details of
these systems since we are here interested mainly in their
algebraic description and in the matching  with the quantum
oscillator in the continuous limit.

The advantage with respect to the previous $SU(2)$ case is now
that the non-compactness of $SU(1,1)$ guarantees that only the
matrix elements of the rising and lowering operators are modified
in the contraction procedure. Since the $SU(1,1)$ group is well
known (see e.g. \cite{PER}), we only recall that it is locally
isomorphic to the (proper) Lorentz group in two spatial dimensions
$SO(2,1)$ and it differs from $SU(2)$ only in a sign in the
commutation relation: $[L_+, L_-] = - 2 L_3$. $SU(1,1)$
representations are well known, in particular the discrete series
$D^+_k$ is
\bea
L_3 |n\ran &=& (n+k)|n\ran\non\,, \\
L_+ |n\ran &=& \sqrt{(n+2k)(n+1)} |n+1\ran \,, \non\\
L_- |n\ran &=& \sqrt{(n+2k-1)n} |n-1\ran\,,\label{su11}
\eea
where, like in $h(1)$, $n$ is any integer greater or equal to zero
and the highest weight $k$ is a non-zero positive integer of
half-integer number.

In order to study the connection with the quantum harmonic
oscillator, we set
\bea
H/\omega = L_3-k+1/2 ~.  \\
a^\dag = L_+/\sqrt{2k} ~,~~~~ a = L_-/\sqrt{2k}.
\eea
The $SU(1,1)$ contraction $k \to \infty$  again recovers the
quantum oscillator Eqs.(\ref{h1}), (\ref{ca}), i.e. the $h(1)$
algebra.  From (\ref{su11}), as announced, we see that the
contraction $k \to \infty$ does not modify $L_3$ and its spectrum
but only the matrix elements of $L_\pm$. The relevant point is
that, while in the $SU(2)$ case the Hilbert space gets modified in
the contraction limit, in the present $SU(1,1)$ case the Hilbert
space is not modified in such a limit: a mathematically well
founded perturbation theory can be now formulated (starting from
Eqs.(\ref{su11}), with perturbation parameter $\propto {1/k}$) in
order to recover the wanted Eqs.(\ref{h1}) in the contraction
limit.

\section{The zero point energy}

We now concentrate on the phase factor in Eq.(\ref{2}), which
fixes the zero point energy in the oscillator spectrum. It is well
known that the zero point energy is the true signature of
quantization and is a direct consequence of the non-zero
commutator of $\hat x$ and $\hat p$. Thus this is a crucial point
in the present analysis.

The $SU(2)$ model considered in Section II  says nothing about the
inclusion of the phase factor.

On the other hand, it is remarkable that the  $SU(1,1)$ setting,
with $H =\om L_3$, always implies a non-vanishing phase, since $k
> 0$. In particular, the fundamental representation has $k = 1/2$
and thus
\bea
L_3 |n\ran &=& (n+1/2)|n\ran\,,\non \\
L_+ |n\ran &=& (n+1) |n+1\ran\,, \non \\
L_- |n\ran &=& n |n-1\ran\, . \label{ho}
\eea
We note that the rising and lowering operator matrix elements do
not carry the square roots, as on the contrary  happens for $h(1)$
(cf. e.g. Eqs.(\ref{h1})).

Then we introduce the following mapping in the universal
enveloping algebra of $su(1,1)$:
\bea \lab{holstein}
&& a= \frac{1}{\sqrt{L_3 + 1/2}} \, L_-
\quad ; \quad
a^\dag= L_+ \, \frac{1}{\sqrt{L_3 + 1/2}}
\eea
which gives us the wanted $h(1)$ structure of Eq.(\ref{h1}), with
$H =\om L_3$. Note that now no limit (contraction) is necessary,
i.e. we find a one-to-one (non-linear) mapping between the
deterministic $SU(1,1)$ system and the quantum harmonic
oscillator. The reader may recognize the mapping
Eq.(\ref{holstein}) as the non-compact analog \cite{gerry} of the
well-known Holstein-Primakoff representation for $SU(2)$ spin
systems \cite{holstein,Shah}.

We remark that the $1/2$ term in the $L_3$ eigenvalues now is
implied by the used representation. Moreover, after a period $T =
2\pi/\omega$, the evolution of the state presents a phase $\pi$
that it is not of dynamical origin: $e^{-iHT} \neq 1$, it is a
geometric-like phase (remarkably, related to the isomorphism
between $SO(2,1)$ and $SU(1,1)/Z_2$ ~($e^{i 2\times 2\pi L_3} =
1$)). Thus the zero point energy is strictly related to this
geometric-like phase (which confirms the result of Ref.
\cite{BJV}).

\section{The dissipation connection}

Eqs.(\ref{su11}) and (\ref{ho})  suggest to us one more scenario
where we may recover the already known connection \cite{thof3,BJV}
between dissipation and quantization. Indeed, by introducing the
Schwinger-like two mode $SU(1,1)$ realization in terms of
$h(1)\otimes h(1)$, the square roots in the eigenvalues of $L_+$
and $L_-$ in Eq.(\ref{ho}) may also be recovered. We set:
\bea\non
&&L_+ \equiv
{A^{\dagger}}B^{\dagger} \quad , \quad L_- \equiv AB \equiv
{L_+}^{\dagger},
\\
&&L_3 \equiv {\frac{1}{2}}(A^{\dagger}A +
B^{\dagger}B + 1) ,
\eea
with $[A, A^{\dagger}] = [B, B^{\dagger}]= 1$ and all other
commutators equal to zero. The Casimir operator is ${\cal C}^{2} =
1/4 + L_{3}^{2} - 1/2(L_{+}L_{-} + L_{-}L_{+})  = 1/4(A^{\dagger}A
- B^{\dagger}B )^{2}$.

We now denote by $\{|n_{A},n_{B}\ran\}$ the set of simultaneous
eigenvectors of the $A^{\dagger}A$ and  $B^{\dagger}B$ operators
with $n_{A}$, $n_{B}$ non-negative integers. We may then express
the states $|n\ran$ in terms of the basis $|j,m\ran$, with $j$
integer or half-integer and $m \ge |j|$, and
\bea
{\cal C}|j,m\ran &=& j|j,m\ran, ~~~~~~~j = 1/2(n_{A} - n_{B}) ~,\\
L_{3}|j,m\ran &=& (m + 1/2)|j,m\ran, ~~~m = 1/2(n_{A} + n_{B}) ~,
\eea
where $n = m - |j|$ and $k = |j| + 1/2$ (cf. Eq.(\ref{su11})).
Clearly, for $j = 0$, i.e. $n = n_A = n_B$, we have the
fundamental representation (\ref{ho}) and $L_{-}|n\ran= AB|n\ran =
{\sqrt n}{\sqrt n}|n\ran = n|n\ran$ (and similarly for $L_+$).
This accounts for the absence of square roots in Eqs.(\ref{ho}).

In order to clarify the underlying physics, it is convenient to
change basis: $|\phi_{j,m}\ran \equiv e^{{\frac{\pi}{2}}L_1}
|j,m\ran$. By exploiting the relation \cite{CRV}
\bea\lab{19a}
i ~e^{{\frac{\pi}{2}}L_1}~L_{3}~e^{-{\frac{\pi}{2}}L_1}~~ =~~ L_{2}~,
\eea
we have
\bea
\lab{19}
L_{2}~|\phi_{j,m}\ran ~=~ i ~(m + 1/2) ~|\phi_{j,m}\ran ~.
\eea
Here it is necessary to remark that one should be careful in
handling the relation (\ref{19a}) and the states
$|\phi_{j,m}\ran$. In fact Eq.(\ref{19a}) is a non-unitary
transformation in  $SU(1,1)$ and  the states $|\phi_{j,m}\ran$  do
not provide a unitary irreducible representation (UIR). They are
indeed not normalizable states \cite{FT,AGI} (in any UIR of
$SU(1,1)$, $L_{2}$ should have a purely continuous and real
spectrum \cite{lindblad}, which we do not consider in the present
case). It has been shown that these pathologies can be amended by
introducing a suitable inner product in the state space
\cite{CRV,BJ,FT} and by operating in the Quantum Field Theory
framework.

In the present case, we set the Hamiltonian to be
\bea
\lab{20}
H &=& H_{0} + H_{I}~. \\
\non
H_{0} &=& \Omega (A^{\dagger}A - B^{\dagger}B) = 2\Omega {\cal C}\, ,\\
H_{I} &=& i\Ga(A^{\dagger}B^{\dagger} - AB) = - 2 \Ga L_{2}\,.
\eea
Here we have also added the constant term $H_{0}$ and set
$2 \Ga\equiv\om$.

In Ref.\cite{CRV} it has been shown that the Hamiltonian
(\ref{20}) arises in the quantization procedure of the damped
harmonic oscillator. On the other hand, in Ref.\cite{BJV}, it was
shown that the above system belongs to the class of deterministic
quantum systems \`a la 't Hooft, i.e. those systems who remain
deterministic even when described by means of Hilbert space
techniques. The quantum harmonic oscillator emerges from the above
(dissipative) system when one imposes a constraint on the Hilbert
space, of the form $L_2|\psi\ran = 0$. Further details on this may
be found in Ref.\cite{BJV}.

\section{Conclusions}

\begin{figure}[t]
%\vspace{-1cm}
\centerline{\hspace{0.6cm}\epsfysize=2.8truein
\epsfbox{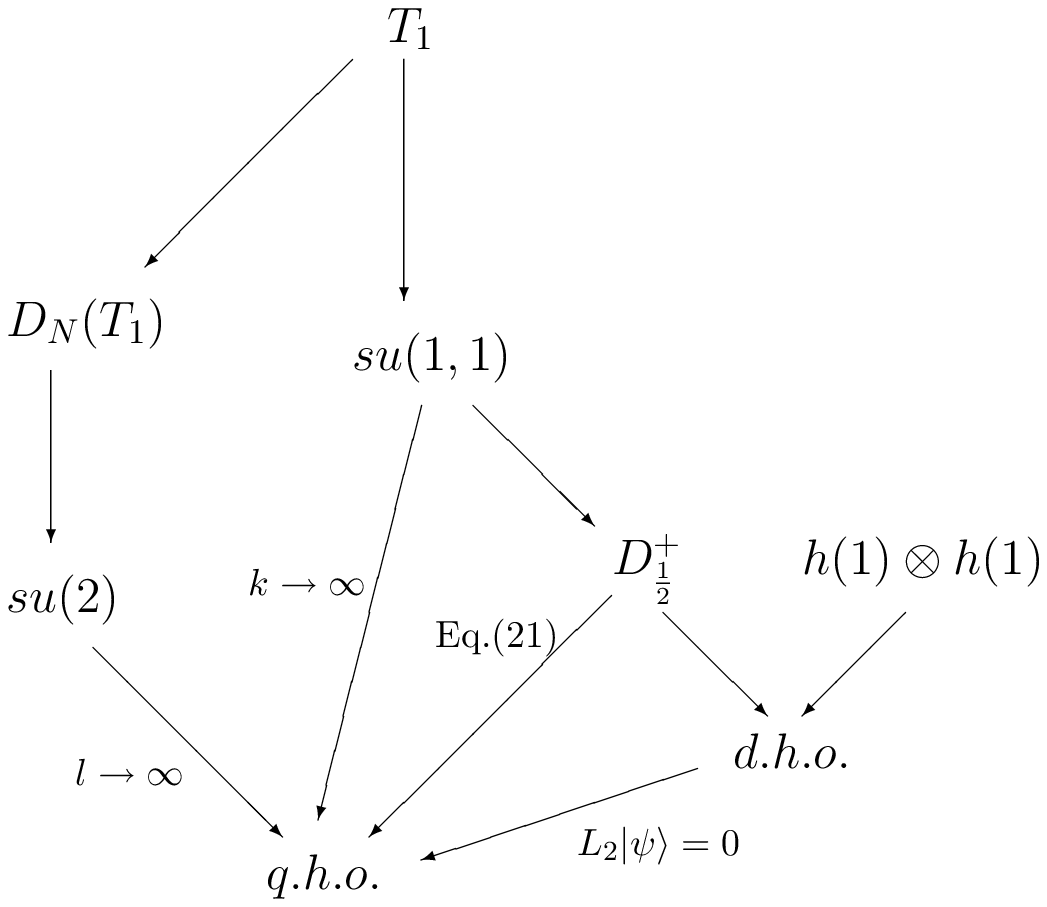}} \vspace{0.5cm} \caption{A schematic
representation of the different quantization routes explored in
this paper. The left route represent 't Hooft procedure, with
contraction of $su(2)$ to $h(1)$.}

\vspace{0.3cm} \hrule
\end{figure}

In this paper, we have discussed algebraic structures underlying
 the quantization  procedure recently proposed by G.'t Hooft
\cite{thof1,thof3}. We have shown that the limiting procedure used
there for obtaining truly quantum systems out of deterministic
ones, has a very precise meaning as a group contraction from
$SU(2)$ to the harmonic oscillator algebra $h(1)$.

We have then explored the r\^ole of the non-compact group
$SU(1,1)$ and shown how to realize the group contraction to $h(1)$
in such case. One advantage of working with $SU(1,1)$ is that its
representation Hilbert space is infinite dimensional, thus it does
not change dimension in the contraction limit, as it happens for
the $SU(2)$ case.

However, the most important feature appears when we consider the
$D^+_k$ representations of $SU(1,1)$, and in particular
$D^+_{1/2}$: we have shown that in this case the zero-point energy
is provided in a natural way with the choice of the
representation. Also, we realize a one-to-one mapping of the
deterministic system onto the quantum harmonic oscillator. Such a
mapping is an analog of the well known Holstein-Primakoff mapping
used for diagonalizing  the ferromagnet Hamiltonian
\cite{holstein,Shah}.

Finally, we have given a realization of the $SU(1,1)$ structure in
terms of a system of damped-amplified oscillators \cite{CRV} and
made connection with recent results \cite{BJV}.

\section*{Acknowledgments}
We acknowledge  the ESF Program COSLAB, EPSRC, INFN and INFM for
partial financial support.

\end{document}